\documentclass{epl}

\title{Exactly solvable model of $A + A \to 0$ reactions on a 
heterogeneous catalytic chain.}
\shorttitle{Reactions on a random catalytic chain}
\author{G.Oshanin\inst{1}
\and O.B\'enichou\inst{2}
\and A.Blumen\inst{3}}
\institute{
  \inst{1}  Laboratoire de
Physique Th\'{e}orique des Liquides, Universit\'{e} Paris VI, 4 Place
Jussieu, 75252 Paris Cedex 05, France\\
  \inst{2}Laboratoire de Physique de la Mati{\`e}re Condens{\'e}e,
Coll{\`e}ge de France, 11 Place M.Berthelot, 75252 Paris Cedex 05,
France\\
  \inst{3} Theoretische Polymerphysik, Universit{\"a}t Freiburg, Hermann-Herder-Str. 3,\\
  D-79104 Freiburg, Germany }

\pacs{68.43.De}{Statistical mechanics of adsorbates}
\pacs{82.65.+r}{Surface and interface chemistry; heterogeneous catalysis at surfaces}

\begin{document}

\maketitle

\begin{abstract}
We present an exact solution describing equilibrium properties
of the catalytically-activated $A + A \to 0$
reaction 
taking place on a 
one-dimensional lattice, where  some of the sites
possess special "catalytic" properties.
The $A$ 
particles undergo continuous 
exchanges with the vapor phase; 
two 
neighboring adsorbed 
$A$s react 
when
at least one of 
them resides on a catalytic site (CS). 
We consider three 
situations for
the CS 
distribution:
regular, $annealed$ random and 
$quenched$ random. 
For all three  
CS distribution types,
we derive 
$exact$ results for the 
disorder-averaged pressure and 
present exact asymptotic 
expressions for 
the particles' mean density.
\end{abstract}

\section{Introduction}

Catalytically-activated reactions (CARs), 
i.e. 
reactions involving
 particles 
which may recombine only
in the presence of some third substance - 
a catalyst - 
are widespread
in nature  ~\cite{1}. 
The properties of such reactions
have attracted
considerable interest 
recently, 
following
an early analysis of a 
particular
reaction scheme, namely the 
CO-oxidation in the presence of 
metal surfaces with  
catalytic properties 
~\cite{3a}. 
It has been realized that 
this reaction exhibits an
essentially different behavior when compared to the
predictions of the classical, 
formal kinetic 
scheme ~\cite{1}
and that
under certain conditions
such
collective phenomena 
as phase transitions
or the formation of
bifurcation 
patterns may occur  ~\cite{3a,4a,rev}.

A common assumption of these studies ~\cite{3a,4a,rev} is that the 
catalyst is modelled
as an ideal surface 
with homogeneous catalytic properties.
On the other hand, in realistic systems, 
the catalyst is often not  a
well-defined object, 
but rather consists of  mobile or localized
 catalytic
sites (CSs) or islands, 
whose spatial distribution 
is complex ~\cite{1}.
Metallic catalysts, for instance, 
are often disordered
compact aggregates, the building blocks 
of which are imperfect crystallites
with broken facets, kinks and steps.
In porous materials with convoluted surfaces,
such as e.g.,
silica, alumina or carbons, 
the effective catalyst 
occupies only a portion of the total surface; 
in solution,
the 
catalyst can consist of active
groups attached 
to polymer chains.

Such complex morphologies render the 
theoretical analysis difficult and, 
as yet, only empirical approaches have been proposed 
(see, e.g. Ref.\cite{1}). 
Consequently, 
analytical solutions even of
idealized or simplified 
models
are 
highly desirable, since such studies may
  provide an understanding 
of the  effects of heterogeneities 
on 
the properties of CARs.

In this Letter we study  the equilibrium  properties  of  a 
 heterogeneous, catalytically-activated 
$A + A \to 0$ reaction 
in  a simple, one-dimensional
(1D) model.
The catalyst is modelled here as an array of special catalytic sites (CSs).
In regard to the CS distribution, we focus 
on three 
situations: regular, $annealed$ random and 
$quenched$ random. 
For all three cases
we derive 
$exact$ results, 
taking into  account  equilibrium fluctuations, 
for the disorder-averaged pressure
and 
present exact asymptotic expressions for 
the particles' mean density.  We show that 
despite the apparent simplicity of the model, one obtains
highly non-trivial behaviors.

\section{Model}

Consider a 1D regular lattice containing $N$ adsorption sites (Fig.1),
which is
brought in contact with a reservoir (vapor phase) 
of non-interacting 
$A$ particles
with hard-cores, maintained at a constant chemical potential $\mu$; the  
activity is hence $z = \exp(\beta \mu)$.

\begin{figure}
\onefigure{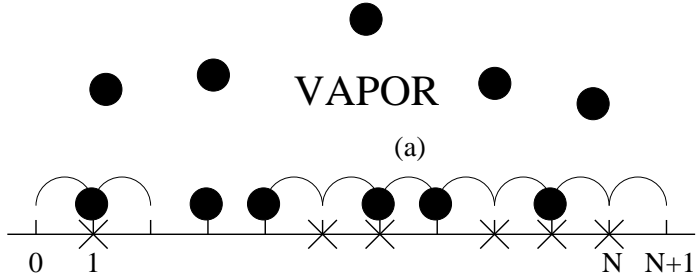}
\caption{1D lattice of $N$ adsorption sites in contact with the
vapor phase. The filled circles denote $A$ particles. The crosses denote the
adsorption sites with catalytic properties (CS). (a) denotes a "forbidden" particle
configuration.}
\label{Fig1}
\end{figure}

The $A$ particles from
the vapor phase can adsorb onto vacant adsorption sites
and desorb back to the reservoir. The occupation 
of the "i"-th adsorption site is
described by the variable $n_i$ such 
that $n_i = 1$ if the "i"-th site is occupied, and $n_i = 0$ otherwise.
We assume that some of the adsorption sites possess 
"catalytic" properties (denoted by crosses in Fig.1)
in the
sense that they induce an immediate
 reaction $A + A \to 0$ between neighboring $A$ particles; that
is, if at least one of two neighboring adsorbed $A$ particles resides on a
catalytic site, these two particles instantaneously react and
 leave the chain. In case when both neighboring sites are occupied,
the $A$ particle landing on the CS reacts with either of its neighbors
with equal probability.
The positions 
of the catalytic
sites are specified by the variable
$\zeta_i$, such that $\zeta_i = 1$ if the "i"-th site is catalytic, and $\zeta_i = 0$ otherwise. 
For convenience, we 
add two boundary sites, i.e. $i = 0$ and $i = N + 1$, and
stipulate that these sites are always non-catalytic,
i.e. $\zeta_0 = \zeta_{N + 1} = 0$, 
and always unoccupied, $n_0 =
n_{N + 1} = 0$.

We  focus  here on  the  equilibrium  properties  of the  model  under  study. 
For a given CS distribution, the
partition function $Z_N(\zeta)$ of
the  adsorbate,  being  in 
equilibrium  with  the  reservoir  and  constrained  locally 
by  the  condition  that  no 
two   $A$  particles 
can  occupy  simultaneously   
two  neighboring  sites  if  at  least  one  of  them  is  a  CS,  
can be written as:
\begin{equation}
\label{partition2}
Z_N(\zeta) = \sum_{\{n_i\}} \prod_{i=1}^N z^{ n_i}
 \Big(1 - \zeta_i n_i n_{i-1}\Big)
\Big(1 - \zeta_i n_i n_{i+1}\Big).
\end{equation}
In this way, any two neighboring sites  $i$ and $i-1$ are
coupled by the factor
$(1 - n_i n_{i-1})$ when at least one of these sites is catalytic.
These coupling
factors are
depicted in Fig.1 as arcs connecting  neighboring sites; note that the
configurations $\{n_i\}$ in 
which the variables
$n_i$ and $n_{i-1}$ 
of two connected sites 
assume both the value $1$ are excluded. 
Note  also  that  the  partition  function  in eq.(\ref{partition2})      
involves  effectively  three-site  interactions.
We introduce now the notion of  a "$K$-cluster", as being a set
of $K$ sites connected to each other
by arcs.
Note that the boundary between adjacent clusters is given by a pair
of two
neighboring, non-catalytic sites, i.e. when 
two consecutive variables $\zeta_i = \zeta_{i+1} \equiv 0$
(see Fig.1).
In this case
the chain decomposes into disjunct clusters and 
$Z_N(\zeta)$ factorizes into independent terms.

We finally  remark  that  for $\zeta \equiv 1$,  $Z_N(\zeta)$  in  eq.(\ref{partition2})  can  be  thought  
of  as  a  1D  version  of  models 
describing  
adsorption  of  hard-molecules, such as, e.g.,
the  so-called  "hard-squares"  ~\cite{fisher}  or the "hard-hexagons" 
models  ~\cite{baxter}. 

\section{Periodic CS distribution}

Consider first a situation with a periodic, (with
period $L$), CS distribution, such that
$\zeta_i = \delta(i,n L + 1)$ with $n = 0,1, \ldots, max(n)$, 
where 
$\delta(k,m)$ is the Kroneker-delta symbol, $\delta(k,m) = 1$ if $k = m$ and $\delta(k,m) = 0$ otherwise,
$max(n) = [(N - 1)/L]$ and
$[x]$ denotes the integer part of
the number $x$.
In this case, the CS density is $p = 1/L$.

We have now to distinguish between two situations: namely, when $L \geq 3$ and when $L
= 1$ or $L= 2$. In the former case, evidently, the factors $(1 - \zeta_i n_i n_{i \pm 1})$
in eq.(\ref{partition2})
are non-overlapping; $Z_N(\zeta)$ decomposes into
elementary three-clusters centered around each catalytic site and (possibly) into
uncoupled, "free" sites,
unaffected by any of the factors $(1 - \zeta_i n_i n_{i \pm 1})$.
On the other side,
for $L = 1$ or $L = 2$ 
we deal with a single cluster spanning the
entire chain. In fact, the role of $L = 1$ and $L = 2$ is,
chemically speaking, identical.

In the case $L \geq 3$ the partition function in
eq.(\ref{partition2})
decomposes into the product
\begin{equation}
\label{deco}
Z_N(\zeta) =
Z_3^{{\cal N}_3} \;  Z_2^{{\cal N}_2} \; Z_1^{{\cal N}_1},
\end{equation}
where $Z_K$, $(K = 1, 2, 3$),
are the partition functions for $1$-, $2$-  and $3$-clusters, respectively:
\begin{eqnarray}
\label{rec1}
Z_1 &=& \sum_{\{n_1 = 0,1\}} z^{n_1} = (1 + z); \;\;\; 
Z_2 =
\sum_{\{n_1,n_2 = 0,1\}} z^{n_1 + n_2} (1 - n_1 n_2)
=  (1 + 2 z); \nonumber\\
Z_3 &=& \sum_{\{n_1,n_2,n_3 = 0,1\}} z^{n_1 + n_2 + n_3} (1 - n_1 n_2) (1 - n_2 n_3)
=  (1 + 3 z + z^2),
\end{eqnarray}
while ${\cal N}_K$
stand for the numbers of such clusters in the
$N$-chain. 
Noticing next that ${\cal N}_3 = max(n) - \delta(max(n),(N - 1)/L)$, 
${\cal N}_2 = 1 + \delta(max(n),(N - 1)/L)$ and using the conservation law
$3 {\cal N}_3 + 2 {\cal N}_2 + {\cal N}_1 \equiv N$, we find that here 
the pressure per site
is given
by
\begin{equation}
\label{free}
 \beta P^{(reg)}(L) =  \lim_{N \to \infty} \Big(\ln Z_N(\zeta)/N\Big) 
=  p \; \ln(1 + 3 z + z^2) + (1 - 3 p) \; \ln(1 + z), \;\;\; (L \geq 3).
\end{equation}
Now, in the periodic case with $L = 1$ or $L = 2$ one finds that 
the partition function of an $N$-site chain obeys the 
three-term  recursion,
$Z_N =  Z_{N - 1} +
z Z_{N - 2}$,
whose first three terms  are given by eqs.(\ref{rec1}). Consequently, 
the $Z_N$ are polynomial functions of the activity $z$, so that: 
\begin{equation}
\label{fi}
Z_N =  \sum_{l =0}^{[(N+1)/2]} {N-l+1
\choose
l}    z^l = z^{(N+1)/2} F_{N + 2}(1/\sqrt{z}),
\end{equation}
where $F_{N}(x)$ are the Fibonacci polynomials \cite{fibo} and ${N
\choose
l}$ denote the binomial coefficients.  In this  case, the pressure 
$P^{(reg)}$ per site for $L = 1$ or $L = 2$ 
turns out to be:
\begin{equation}
\label{i}
\beta P^{(reg)} =  \ln\Big(\frac{\sqrt{1 + 4 z} + 1}{2}\Big), \;\;\; (L = 1 \;\;\; {\rm or} \;\;\; L = 2).
\end{equation}
By differentiating the results in
eqs.(\ref{free}) and (\ref{i}) with respect to the chemical potential $\mu$,
we find that in the asymptotic limit $z \to \infty$
the mean density $n^{(reg)}$ 
of adsorbed 
particles is
\begin{equation}
\label{asas}
n^{(reg)}(p) = \left\{\begin{array}{ll}
\displaystyle (1 - p) - \frac{1}{z} + {\cal O}\Big(\frac{1}{z^2}\Big),  \;\;\;   \mbox{$L \geq 3$,} \nonumber\\
\displaystyle \frac{1}{2} - \frac{1}{4 \sqrt{z}} + {\cal O}\Big(\frac{1}{z^{3/2}}\Big),  \;\;\;   \mbox{$L = 1 \;\;\; {\rm and} \;\;\; L = 2$.}
\end{array}
\right.
\end{equation}
This signifies that 
for $L = 1$ or $L = 2$ and $z \to \infty$ the system undergoes an ordering transition.

\section{Random CS distribution. Annealed disorder}

In this case 
the disorder-average pressure $P^{(ann)}(p)$
per site obeys $\beta P^{(ann)}(p) = 
\lim_{N \to \infty} (\ln <Z_N(\zeta)>/N)$ and our aim is to evaluate $<Z_N(\zeta)>$.
Averaging $Z_N(\zeta)$ in eq.(\ref{partition2}), 
we obtain
\begin{equation}
\label{zu}
 <Z_N(\zeta)> = \sum_{\{n_i\}}
\left(z^{\sum_{i=1}^N n_i}\right) \left( (1 - p)^{\sum_{i=1}^N  \Psi_i}\right),
\end{equation} 
where $\Psi_i$ is the 
three-site indicator function of the form $\Psi_i = (n_i n_{i+1} + n_i n_{i-1} -
n_{i-1} n_i n_{i+1})$.  
Note now that $\Psi_i$ always equals zero
for unoccupied sites, $(n_i = 0)$,
and assumes 
the value $\Psi_i
= 1$ only for the occupied sites, $(n_i = 1)$, 
which have at least one (or two)
occupied neighboring sites. Consequently, one has that
$\sum_{i=1}^N \Psi_i = {\cal N}_+[\{n_i\}]  - {\cal N}_{is}[\{n_i\}]$,
where ${\cal N}_+[\{n_i\}]$ is the number of lattice
sites on which  (in a given realization
$\{n_i\}$) the occupation variable $n_i$ 
assumes the value $1$, while ${\cal
N}_{is}[\{n_i\}]$
is the
realization-dependent number of 
isolated occupied sites (elementary
cells of the form $(0,1,0)$).
Hence, $<Z_N(\zeta)>$ in eq.(\ref{zu})
can be rewritten as 
\begin{equation}
\label{gen}
<Z_N(\zeta)> = \sum_{{\it N}_+ =0}^{N}
\Big(z (1 - p)\Big)^{{\it N}_+} \; \sum_{m = 0}^{N - {\it N}_+ +1}
(1 - p)^{-m} M_m({\it
N}_+|N),
\end{equation}
where
$M_m({\it
N}_+|N)$ stands for the number of
realizations $\{n_i\}$
that have a fixed ${\it N}_+$
and contain $\it exactly$
$m$ elementary cells $(0,1,0)$. Using  combinatorial arguments, we get ~\cite{mi}:
\begin{equation}
\label{y}
M_m({\it
N}_+|N) = \frac{1}{2 \pi i} \; \frac{({\it N}_{-}+1)!}{ m! ({\it N}_{-}+1 - m)!} \oint_{{\cal C}} \; \frac{d \tau}{\tau} \;
\tau^{\displaystyle \Big({\it N}_+ - m\Big)} \; \left(\frac{1 - \tau + \tau^2}{1 - \tau} \right)^{N - {\it N}_+ + 1 -m},
\end{equation}
where ${\cal C}$ stands for any closed contour which encircles the origin 
counterclockwise. 
Substituting $M_m({\it
N}_+|N)$, 
eq.(\ref{y}), into eq.(\ref{gen}), and 
performing the summations, we are able to 
determine $<Z_N(\zeta)>$, and hence, also $P^{(ann)}(p)$; 
in the thermodynamic limit  
$P^{(ann)}(p)$ obeys:
\begin{equation}
\label{anni}
\beta P^{(ann)}(p) = \ln\left(3 z (1-p)/\left[1 - 6 \sqrt{Q} \sin\Big( \displaystyle \frac{1}{3}
\arcsin\Big(R/\sqrt{Q^3}\Big)\Big)\right]\right),
\end{equation}
where
\begin{equation}
R = \frac{1}{27} +  \frac{1}{6} \Big(  \frac{(1-p)}{p}(1 + z (1-p)) -
\frac{3 z (1-p)^2}{p} \Big); \;\;\; Q = \frac{1}{9} + \frac{(1-p)}{3 p}(1 + z (1-p)).
\end{equation}

Next, differentiating eq.(\ref{anni}) with respect to $\mu$, one finds
the particles' mean  density $n^{(ann)}(p)$.
The resulting expression is rather 
cumbersome and we present it elsewhere ~\cite{mi}; 
here we merely display 
the asymptotic behavior of $n^{(ann)}(p)$ in the large-$z$ limit.
 
We note that in this limit 
the forms of 
$n^{(ann)}(p)$ 
for $p < 1$ 
and for $p \equiv 1$ 
are quite different, which 
implies that $p \equiv 1$ is a special point.
For $p < 1$ and  $z \gg (1 - p)^{-2}$, we obtain
\begin{equation}
n^{(ann)}(p) = 1 - \frac{1}{(1 - p) z} + \frac{(1 - 3 p)}{(1 - p)^3 z^2} + {\cal O}\Big(\frac{1}{z^3}\Big),
\end{equation} 
while for $p \equiv 1$ and $z \to \infty $ the particles'
 mean  density 
obeys the expression valid for $L = 1$ or $L = 2$,
given in the 
second line of
eq.(\ref{asas}). Note that
for $p$ arbitrarily close but not equal to unity, $n^{(ann)} \equiv 1$ 
as $z \to \infty$, 
while for $p \equiv 1$ one has $n^{(ann)} \equiv 1/2$ as 
$z \to \infty$.

This behavior can be understood as follows: In the annealed disorder case, 
instead of averaging $\ln Z_N(\zeta)$,
we can average
$Z_N(\zeta)$ itself.
Then, the disorder-averaged pressure is 
defined by the "effective" partition 
function in eq.(\ref{zu}).
Here, a strict constraint that no two particles can occupy 
neighboring sites if at 
least one of them sits on the CS, is replaced
by a more tolerant condition, 
which allows for such pairs to be present
but  a penalty  $\epsilon = 2 \ln(1 - p)$ is to be paid. 
For any 
$p < 1$,  the penalty $|\epsilon| < \infty$ 
is finite and one thus expects 
that
for $p < 1$ and $\beta \mu \gg |\epsilon|$
the leading behavior is that of the 
trivial 
Langmuir adsorption model;
hence, 
$n^{(ann)} \sim z/(1+z) \to 1$ as $z \to \infty$. On the other hand, 
for $p \equiv 1$,  $\epsilon$ becomes infinitely large  
and can not 
be compensated by 
increasing $\beta \mu$.     

\begin{figure}[ht]
\twofigures[width=5cm, angle=270]{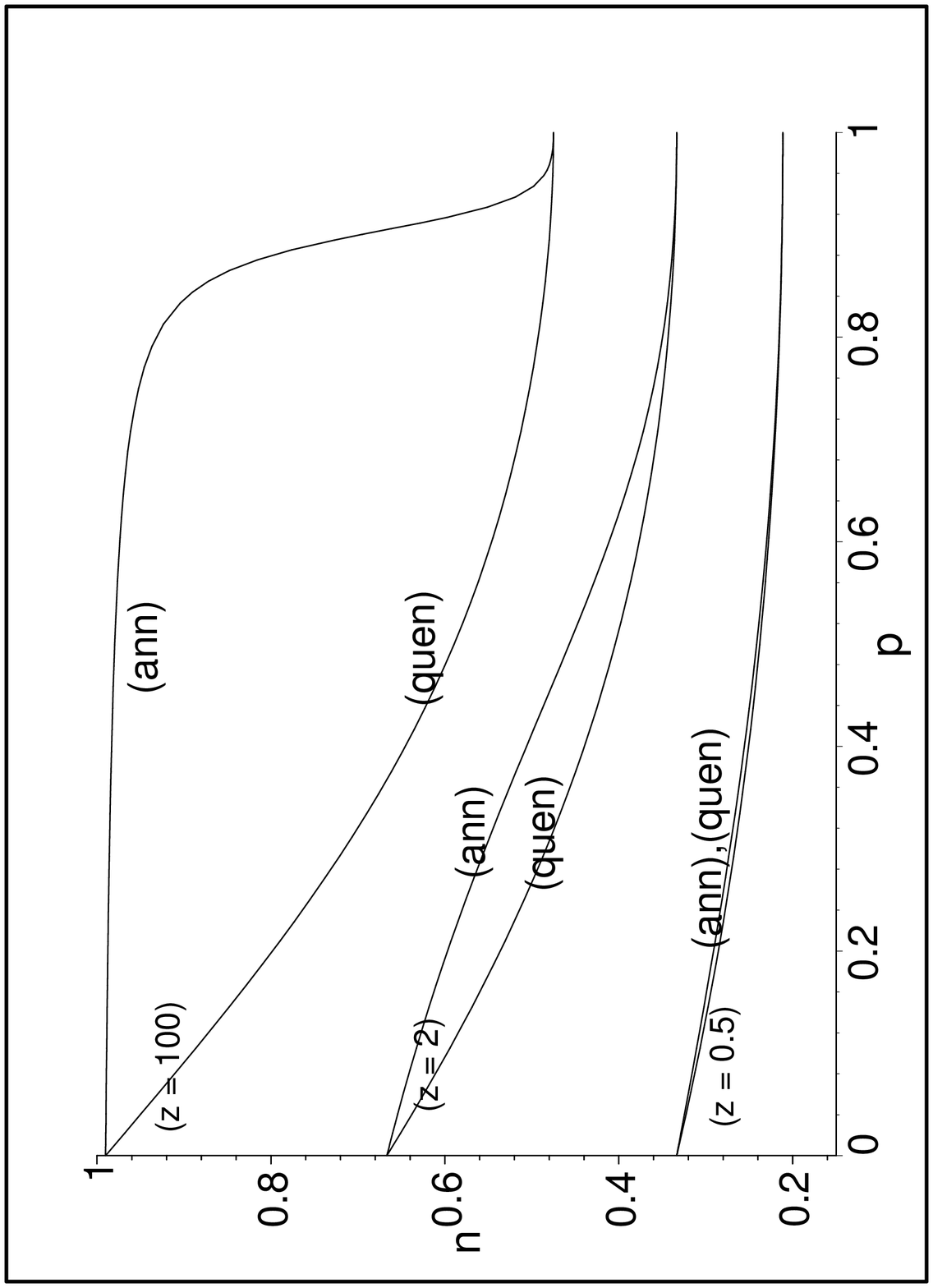}{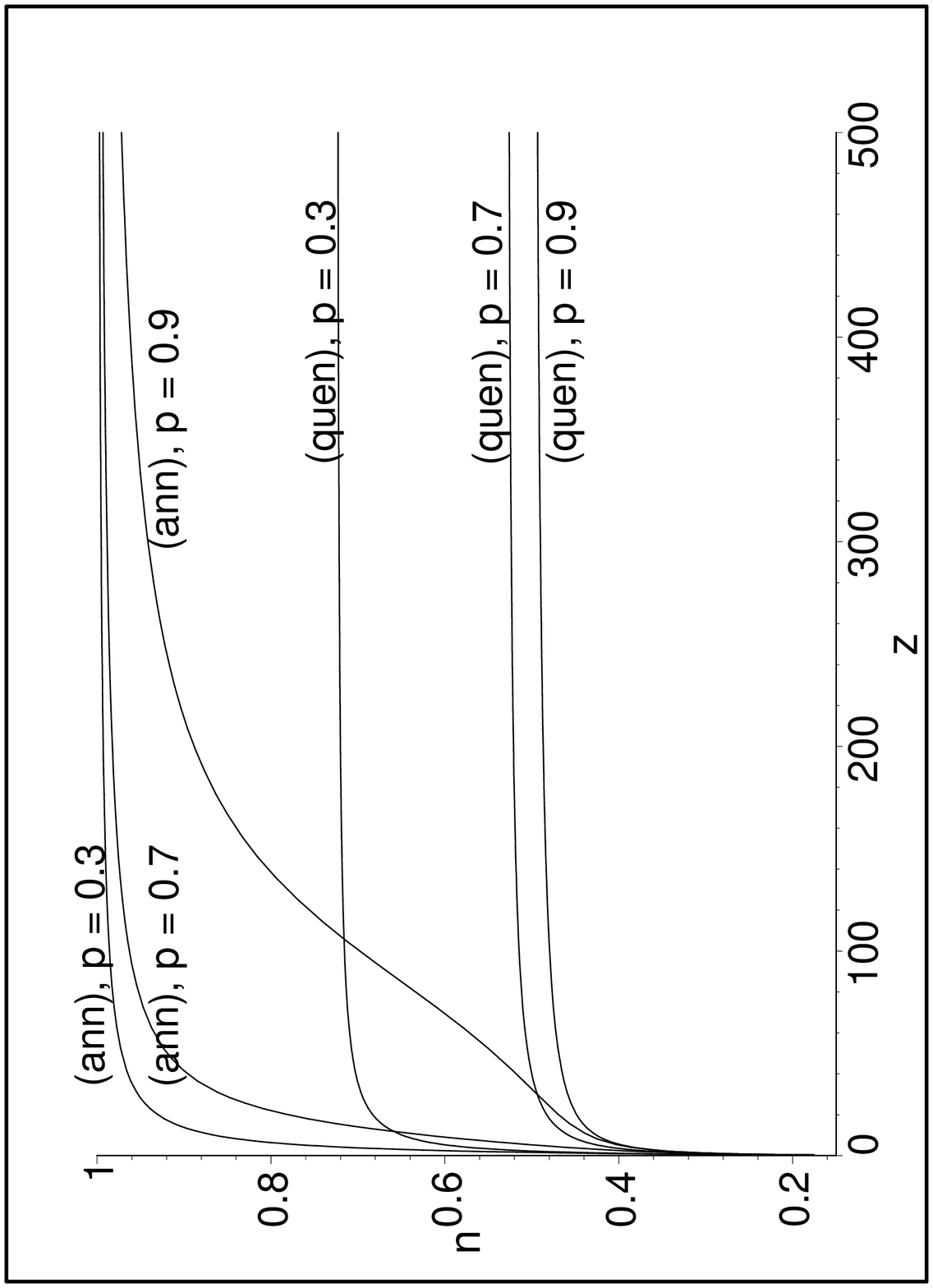}
\caption{Particles' mean density $n$
versus the CSs mean density  $p$ for different values of $z = \exp(\beta \mu)$. Curves with 
the signs (ann) and (quen)  
depict the behavior of $n$ 
in case of 
annealed and quenched random CS distributions, respectively. }
\label{Fig2}
\caption{Particles' mean density $n$ in the annealed and quenched disorder cases
 versus the activity $z$ for several different values of $p$.}
\label{Fig3}
\end{figure}

\section{Random CS distribution. Quenched disorder}

We finally turn to the most challenging situation, namely that of  
quenched random CS placement.
Here we only briefly outline the derivation, 
which is rather lengthy and will be presented in detail elsewhere ~\cite{mi}. 

Consider an $N$-chain which contains a fixed number $N - N_{nc}$ of
catalytic sites, and hence $N_{nc}$ non-catalytic sites, 
the latter being placed at the positions $\{X_n\}$,
$n = 1, 2, \ldots , N_{nc}$. Next, we introduce a set of $N_{nc} + 1$ intervals
$\{l_{n}\}$ connecting consecutive 
non-catalytic sites, such that $l_n = X_n - X_{n -1}$ (with $X_0 = 0$) and
$l_{N_{nc} + 1} = N + 1 - X_{N_{nc}}$.  
That is, the first interval reaches from the boundary 
site $i = 0$ 
to the first
non-catalytic site, 
closest to it,
the second interval goes from this site to the next non-catalytic site and so on, while
 the interval $l_{N_{nc} + 1}$ goes from 
the last non-catalytic site of the chain to the
site $i = N + 1$.

Now, 
the logarithm of the 
partition function in eq.(\ref{partition2}), 
averaged over all realizations of the quenched
random variable $\{\zeta_i\}$, can be formally written as
\begin{equation}
\label{ququq}
<  \ln Z_N(\zeta)> \; = \; \sum_{N_{nc} = 0}^N p^{N - N_{nc}} 
(1 - p)^{N_{nc}} \sum_{\{l_{n}\}}
\ln Z_N(\{l_{n}\}),
\end{equation}
where  the inner sum with the sign
$\{l_{n}\}$ 
extends 
over all possible solutions of the equation
\begin{equation}
l_1 + l_2 + l_3 + \ldots + l_{N_{nc} + 1} = N + 1, \;\;\; l_i \geq 1,
\end{equation}
while $Z_N(\{l_{n}\})$ stands for
the partition function corresponding
to a given set $\{l_{n}\}$ of intervals. 
For each set of intervals 
$Z_N(\{l_{n}\})$ 
decomposes into smaller clusters,
\begin{equation}
\label{deco}
Z_N(\{l_{n}\}) = Z_1^{{\cal N}_1(\{l_{n}\}|N)} \;
Z_2^{{\cal N}_2(\{l_{n}\}|N)} \; Z_3^{{\cal N}_3(\{l_{n}\}|N)} \; \ldots \;
Z_N^{{\cal N}_N(\{l_{n}\}|N)},
\end{equation}
where $Z_K$, $(K = 1, 2, \ldots ,N)$, is the $K$-cluster partition function, 
which obeys
eq.(\ref{fi}) 
(with $N$ replaced by $K$), while
${\cal N}_K(\{l_{n}\}|N)$ denotes the 
(realization-dependent) number of
$K$-clusters in the $N$-chain with $N_{nc}$ non-catalytic sites. 
Taking the logarithm of both sides of eq.(\ref{deco}),
we obtain
\begin{equation}
\label{decom}
\ln Z_N(\{l_{n}\}) = \sum_{K = 1}^N {\cal N}_K(\{l_{n}\}|N) \; \ln Z_K,
\end{equation}
and consequently, $<  \ln Z_N(\zeta)>$ has the form:
\begin{equation}
<  \ln Z_N(\zeta)> \; = \; 
\sum_{K = 1}^N w_{K,N}(p) \ln Z_K,
\end{equation}
where $w_{K,N}(p)$ is the statistical weight
of the $K$-clusters in an $N$-chain:
\begin{equation}
w_{K,N}(p) =  \sum_{N_{nc} = 0}^N p^{N - N_{nc}} \; (1 - p)^{N_{nc}} \; \sum_{\{l_{n}\}} {\cal N}_K(\{l_{n}\}|N).
\end{equation}
The weights $w_{K,N}(p)$ can be also evaluated combinatorially ~\cite{mi}; after some 
straightforward but rather tedious calculations we find that
for $K\neq 1$ and $K\neq N$:
\begin{equation}
\label{w}
w_{K,N}(p)=p^{(K-1)/2} \; (1-p)^{(K+3)/2} \Big\{2F_{K}\left(\phi\right) 
+(1-p)(N-K-1)F_{K-2}\left(\phi\right)\Big\},
\end{equation}
where $\phi = \sqrt{p/(1-p)}$ and
$F_K(\phi)$ are
the Fibonacci polynomials ~\cite{fibo},
while for $K = N$ one has
\begin{equation}
\label{ww}
w_{N,N}(p)=p^{N/2}\;(1-p)^{N/2}\;\Big\{
2F_{N-1}\left(\phi\right)+
\phi F_{N-2}\left(\phi\right)+
\phi F_{N}\left(\phi\right)\Big\}.
\end{equation}
Eventually, we arrive at the following exact expression, which determines  
the disorder-averaged pressure $P^{(quen)}(p)$  per site in the $quenched$ disorder case
 (for arbitrary $p$ and $z$):
\begin{eqnarray}\label{resultat integral}
&&\beta P^{(quen)}(p) = \lim_{N \to \infty} \frac{1}{N} \Big< \ln Z_N(\zeta) \Big> =
\lim_{N \to \infty} \frac{1}{N} \sum_{K = 1}^N w_{K,N}(p) \ln Z_K = (1-p)^3\ln(1+z) + \nonumber\\
&+& p \Big(5 - 7 p + 3 p^2\Big) \ln\left(\frac{\sqrt{1 + 4 z} + 1}{2}\right)
- \frac{p (1 - p)^2}{2} \ln(1 + 4 z) - 
\frac{p(1-p)^4}{\sqrt{p(4-3p)}} \times \nonumber\\
&\times& \sum_{m=0}^\infty\left[\left(\frac{2 p (1 - p)}{\sqrt{p ( 4 - 3 p)} - p}\right)^m
- \left(\frac{- 2 p (1 - p)}{\sqrt{p ( 4 - 3 p)} + p}\right)^m
\right]
\ln\left(1- 
\left(\frac{1 - \sqrt{1 + 4z}}{1 + \sqrt{1 + 4 z}}\right)^{m+5}\right)
\end{eqnarray}
Differentiating eq.(\ref{resultat integral}) and turning to the large-$z$ limit, we find that 
\begin{equation}
\lim_{z \to \infty} n^{(quen)}(p) = 1 - \frac{p}{1 + p^2}.
\end{equation}
This is different from
the behavior observed in the 
annealed disorder case:
$\lim_{z \to \infty} n^{(ann)}(p) \; \equiv 1$ for any $p < 1$; 
it also
differs from our result for the
periodic CS distribution:
$\lim_{z \to \infty}  n^{(reg)}(L)$ $= 1 - p$ for $p \leq 1/2$ and  
$\lim_{z \to \infty} n^{(reg)}(L = 1 \; {\rm or} \; L = 2) \equiv 1/2$ 
for $p = 1/2$ and $p = 1$. Note that  here, 
distinct from the annealed disorder case, 
$n^{(quen)}(p)$  
does not show any discontinuity 
in the limit $z \to \infty$ and $p \to 1$. 
The behavior of
the particles' mean density in the quenched disorder case is also given in Figs.2 and 3. 

\section{Conclusions}

We have presented here an exact lattice solution describing the equilibrium
properties of the
heterogeneous 
catalytically-activated 
$A + A \to 0$ reaction
in the 
case when the $A$ particles undergo continuous exchanges with a reservoir
and react 
immediately if at least one $A$
of a neighboring
$AA$-pair
 sits on a catalytic site. We have considered
three possible situations for the CSs placement on the 1D 
lattice: regular, $annealed$ 
random and $quenched$ random.
In all three cases we have calculated  the disorder-averaged pressure
of the adsorbate exactly and have presented asymptotic results
for the particles' mean density. 
 Remarkably, at equal mean densities of the CSs,
the asympotic values of the mean
 densities differ in 
all three cases. We close by noting 
that the model studied here 
furnishes another example (see, e.g.,  Refs.~\cite{hil,vul})
of a 1D Ising-type system 
with random multi-site
interactions 
which admits an exact solution. 

\section{Acknowledgment}

The support of the Deutsche Forschungsgemeinschaft
and of the Fonds der Chemischen Industrie is gratefully
acknowledged.

\end{document}